\documentstyle[float,aps,prl]{revtex}                

\begin{document}
\draft

\twocolumn[\hsize\textwidth\columnwidth\hsize\csname@twocolumnfalse\endcsname

\title{
Quantum Breathers in a Nonlinear Lattice
}
\author{
W. Z. Wang$^{\star}$, J. Tinka Gammel, A. R. Bishop,
}
\address{
Theoretical Division,
Los Alamos National Laboratory, Los Alamos, New Mexico 87545, USA
}
\author{
and M. I. Salkola
}
\address{
Department of Physics \& Astronomy, McMaster University,
Hamilton, Ontario L8S 4M1, Canada
}

\date{\today}
\maketitle
\tightenlines

\begin{abstract}
We study nonlinear phonon excitations in a one-dimensional quantum
nonlinear lattice model using numerical exact diagonalization. We find
that multi-phonon bound states exist as eigenstates which are natural
counterparts of breather solutions of classical nonlinear systems. In a
translationally invariant system, these quantum breather states form
particle-like bands and are characterized by a finite correlation
length. The dynamic structure factor has significant intensity for the
breather states, with a corresponding quenching of the neighboring
bands of multi-phonon extended states.
\end{abstract}

\

]

\narrowtext

The paradigms of nonlinearity have provided numerous insights into
condensed matter physics~\cite{bishop}. For example, the dynamics of
nonlinear excitations (such as solitons, polarons, and breathers) are
central to understanding thermodynamic and transport properties of
various low dimensional materials~\cite{alan}. As an important example
of intrinsic dynamic nonlinearity, breather excitations --- spatially
localized and time-periodic waves in the form of bound states of linear
excitations --- have been found or excluded in various nonlinear models
depending on the properties of nonintegrability and
discreteness~\cite{peyrard,phillpot}. In particular, although breathers
are rigorously stable in integrable partial differential equations
(such as the (1+1)-dimensional sine-Gordon or nonlinear Schr\"odinger
equations), they are unstable in {\it nonintegrable continuous}
systems. Recently it has been appreciated, however, that they can be
stablized, not only in integrable \cite{cai,page} but also in
nonintegrable~\cite{takeno} cases, by {\it lattice discreteness} and
sufficiently strong nonlinearity --- in the form of dynamic localized
excitations, the generalization of uncoupled oscillators. For some
models, existence regimes have been rigorously
established~\cite{mackay}. We are left with a central question for
physical problems described by discrete, {\it nonintegrable} {\it
quantum} systems: do nonlinear solutions analogous to the classical
breather exist and, if so, what are their observable signatures? This
question has a long history in terms of biphonon bound states or
resonances~\cite{ruvalds}, and has been studied more recently in Bethe
Ansatz systems~\cite{bethe} as well as semi-classically~\cite{dashen}.
Here, we focus on numerical exact diagonalization of a simple
nonintegrable quantum anharmonic chain. We find that the classical
breather solutions indeed have their quantum counterparts as
eigenstates of the system Hamiltonian with distinctive signatures in
appropriate correlation functions.

Our quantum nonlinear lattice model is
\begin{equation}\label{eq1}
H= {\hbar \Omega} 
  \sum_{ n} \big[ {\mbox{$\hat{p}_n^2$} \over  \mbox{2}}
+ {\mbox{$\eta$} \over \mbox{2}} (\hat{\phi}_n-\hat{\phi}_{n+1})^2
+ w \hat{\phi}_n^2 + v \hat{\phi}_n^4 \big] \, ,
\end{equation}
with periodic boundary conditions, and with $\hat{p}_n$ and $\hat{\phi}_n$
the dimensionless canonical lattice momentum and displacement
operators, $[\hat{\phi}_m,\hat{p}_n]= i\delta_{mn}$.  The coefficient
${\hbar \Omega}$ sets the scale of energy, whereas $\eta$, $w$, and
$v$ are dimensionless parameters. The first two terms describe a
standard harmonic Debye lattice, while the second two terms for
$w,v$$\ge$$0$ describe a single-well $\phi^4$ nonlinearity, which
corresponds to the Taylor expansion of either a local nonlinear
interaction due to anharmonicity, or an effective lattice interaction
arising from, $e.g.$, electron-lattice coupling. Other nonlinear
Hamiltonians can be treated by the same methods described below and can
be expected to contain varieties of breathers. However, the simple
form in Eq.\ (\ref{eq1}) isolates the breather without, $e.g.$,
complications of multiply degenerate ground states. Note that the
class of models represented by Eq.\ (\ref{eq1}) does not conserve
the total number of phonon quanta in contrast with simpler discrete quantum
models where solitons have been discussed\cite{scott}.

There are two limiting cases for which the physics of the model becomes
transparent. First, in the absence of nonlinearity ($v$=0), the model
is trivially solvable. The linear coupling between nearest-neighbor
oscillators in Eq.\ (\ref{eq1}) results in spatially extended
Bloch-wave states of phonons.The $w$ term acts merely as an adjustable
factor controlling the dispersion of phonon bands, and will be kept
small throughout this study. Second, for $v$ ($\sim$$1$) $\gg$ $\eta$,
the $\phi^4$ term plays the dominant role. However, we can again
qualitatively understand many features of the system, including the
appearance of breathers. The nearest-neighbor interaction governed by
$\eta$ will cause the local ``Einstein'' modes to become hybridized,
creating narrow continua of phonon states at each local vibrational
energy. The $\phi^4$ term produces a local repulsive interaction
between two or more quanta if they are located at the same site. This
interaction therefore leads to bound multi-phonon states, breathers,
that are split {\it upwards} from each multi-phonon
continuum~\cite{takeno}. These states are the simplest form of
breathers we expect to find in molecular systems, {\it i.e.}, whenever
local nonlinear vibrations are weakly coupled to neighboring ones.
Below, we consider systems with $\eta$$\sim$$v$$\sim$$1$, which are
more interesting because of their relevance to many strongly
interacting solid-state systems, such as quasi-one-dimensional
polymers. In this case, the nearest-neighbor Debye coupling will
compete with the $\phi^4$ nonlinearity to determine the correlation
length of eigenstates in the system. The main physical effects of
the $\phi^4$ term are: ($i$) opening finite gaps in the many-body
excitation spectrum, which offers the possibility of bound states
within the gaps; and ($ii$) suppression of the effective linear
coupling, facilitating the dynamic localization of certain eigenstates.

We study this problem by numerical exact diagonalization using the
Einstein phonon basis~\cite{einstein}, where the only physical
approximation is the truncation of the infinite Hilbert space. This
restricts our numerical simulations to a parameter region where the
nonlinearity is not too large. To maintain physical correctness and
numerical reliability, we solve the problem for finite chains with
moderately large nonlinearity, systematically studying the effect of
truncation and system size. We adopt a modified~\cite{h-footnote}
Davidson method~\cite{davidson} to perform the exact diagonalization
for the large sparse matrix of the system Hamiltonian, and then a
projective Lanczos method~\cite{loh} to calculate various correlation
functions. Eigenfunction convergence has been tested through the
Weinstein lower bound formula~\cite{davidson}, and the symmetries
($e.g.$, the translational invariance under periodic boundary
conditions) were also checked.

The low-lying eigenspectra for 4- and 8-site nonlinear chains with
representative parameter sets are shown in Fig.\ \ref{fig1}. Several
characteristic features can be noted: First, as anticipated above,
finite gaps are opened due to the effect of the nonlinearity
$v$.  Second, states which originally belong to the same band in the
linear Debye or Einstein lattice now split into several bands. While
the nonlinearity does not allow the ordering of these bands in terms of
bare phonons of the linear system, one can still roughly
distinguish different bands by the bare phonon number distribution
function $\rho_{\alpha}(m)=\langle\alpha|\delta(\sum_n b_n^\dagger
b_n-m)|\alpha\rangle$ and the average number of bare phonons $\langle
M_\alpha\rangle = \sum_m m\rho_\alpha(m)$, where $b_n= (\hat{\phi}_n +i
\hat{p}_n)/\sqrt{2}$, and $|\alpha\rangle$ is an eigenstate of the system
with crystal momentum $q_\alpha$ and energy $\epsilon_\alpha$. (We have
also used this function to determine that our results are not dependent
on the truncation.) In contrast to the Debye lattice, the ground state
is isolated, with a finite gap to the excited eigenstates. The quantity
$\rho_{\alpha}(m)$ for the first band above the ground state shows that
it is mainly formed from one phonon states (peaked at $m$=1). There are
several different kinds of higher excited bands of states, including
ones with small but finite bandwidths. The emergence of these isolated
bands can be readily interpreted as the existence of particle-like
states, where the bandwidth is a measure of the inverse particle mass.
They can also be viewed as bound phonon states. We will argue below
that these narrow, isolated bands of states are quantum breathers.
Note that these breather states have explicit Bloch-wave translational
symmetry with well-defined crystal momenta $q_\alpha$ associated with
their center of mass motion. However, they also exhibit a finite lattice 
displacement correlation length (describing the particle-like coherence 
of breathers), as we now discuss.

To identify breathers in the quantum excitation spectrum, we establish
their main characteristics by examining key correlation functions.
These functions are not only important for diagnostic purposes, but are
also relevant to many measurable quantities. One characteristic of the
classical breather is spatial localization of the envelope. In the
translationally invariant quantum case, the dynamic localization of
breathers can be measured by the corresponding spatial correlation of
the displacements at sites separated by $n$ lattice spacings:
\begin{equation}\label{eq2}
f_\alpha(n-n^\prime,t-t^\prime) =
\langle\alpha|\hat{\phi}_{n^\prime}(t^\prime)\hat{\phi}_{n}(t)|\alpha\rangle .
\end{equation}
For the linear system, 
these correlation functions are readily computed and they in
general have multiple spatial and temporal Fourier components
reflecting the system's phonon modes propagating with various phase
velocities. However, in a nonlinear system, multi-phonon
bound states can appear. In Fig.\ \ref{fig2}, the instantaneous spatial
correlation function $f_{\alpha}(n,0)$ of a typical state in the
isolated bands is illustrated. The spatial correlation is clearly
localized at $n=0$ and exhibits the particle-like nature of these
states, a key feature of a breather. In this strongly nonlinear regime,
the breathers have a small correlation length ($\xi$ is on the order of a
lattice constant). The other states form a ``continuum'' of extended
multi-phonon states which have significant weight even at large $n$,
also illustrated in Fig.~2.

We investigated temporal coherence in the localized state by examining
the Fourier transform $F_{\alpha}(n,\omega)$ of the time-dependent
correlation function $f_{\alpha}(n,t)$,
\FL
\begin{eqnarray}\label{eq3}
F_{\alpha}(n-n^\prime,\omega)
=  \sum_{\beta} \langle\alpha|\hat{\phi}_{n^\prime}|\beta\rangle
\langle\beta|\hat{\phi}_{n}|\alpha\rangle
\delta(\hbar\omega - \epsilon_\beta + \epsilon_\alpha),(3) \nonumber
\end{eqnarray}
calculated using the spectral projection method~\cite{loh}.
$F_\alpha(0,\omega)$ for a typical breather state is shown in
Fig.\ \ref{fig3}. The dynamics of the localized states exhibit a small
number of frequencies with significant amplitudes, reflecting the
anharmonicity of the system. The strongly localized breather here can
be expected to exhibit multiple internal frequencies even
classically~\cite{cai,takeno,freq-note}, especially in the strongly
nonlinear regime. However, this multi-frequency property is {\it not}
distinctive for breathers -- anharmonic phonons and multi-phonon states
exhibit similar signatures.

These quantum breathers do show experimentally observable signatures of
their distinctive local structure and dynamics. In particular, lattice
displacements within a few correlation lengths $\xi$ are strongly
correlated in the breather and therefore can be analyzed by the
density-density correlation function, $ S(r,t)= (1/N) \int dr' \langle
\hat{\rho}(r',0)\hat{\rho}(r'+r,t)\rangle,$ where $\hat{\rho}(r,t)=
\sum_n \delta(r-n{\rm a_0}-\ell\hat{\phi}_n(t))$ is the density
operator, ${\rm a_0}$ is the lattice constant, $N$ is the number of
lattice sites, and $\ell$=$(\hbar/M\Omega)^{1/2}$ sets the scale of
length for lattice displacements. The qualitative behavior of $S(r,t)$
is sensitive to the ratio $r/\xi$. In principle, these correlations can
be probed by neutron scattering~\cite{salkola}, which directly measures
the dynamic structure factor $S(q,\omega)$ (the spatial and temporal
Fourier transform of $S(r,t)$), given at zero temperature by
\begin{eqnarray}\label{eq5}
S(q,\omega) &=& \displaystyle \sum_{\alpha}
 \delta(\hbar\omega-\epsilon_\alpha+\epsilon_0)
   \int dr  e^{-iqr} \int dr'\nonumber\\
&&~~~~~~~~~{1 \over N}\langle 0|\hat{\rho}(r',0)|\alpha \rangle
\langle \alpha|\hat{\rho}(r'+r,0)|0 \rangle,~~~(4)\nonumber
\end{eqnarray}
where $|0 \rangle$ is the ground state.  In the linear case, one would
see a response in $S(q,\omega)$ which traces each multi-phonon band
dispersion, falling off exponentially in $q$ ({\it i.e.}, the quantum
Debye-Waller factor) and decreasing algebraically with $\omega$, with
regular spacing in $\omega$. The result for the nonlinear 4-site system
of Fig.\ \ref{fig1}(a) is illustrated in Fig.\ \ref{fig4}. The elastic
response of the ground state at $\omega$=0 has now gained an
exponentially decaying $q$-dependence which is different from the
harmonic Debye-Waller factor~\cite{salkola}. We also note that in this
zero temperature calculation, there is no zero-frequency contribution
from the breathers~\cite{kerr}. In addition to the expected large
low-energy contributions from the ground state and the first phonon
band, breather excitations are the dominant contribution, whereas those
contributions from the extended multi-phonon ``continua'' are almost
negligible. The sum rule, $\int d\omega \,\, \omega S(q,\omega) =
{\Omega} q^2 \ell^2 /2 $, implies that the quenching of the
extended-state contribution is consistent with breathers forming as
bound states of nearby phonons, and is similar to the transfer
mechanism of the optical oscillator strength to local modes in the
presence of electronic bound states. The center of the
response is shifted to higher $q$ for higher breather bands, consistent
with the fact that higher lying breathers (binding more phonons) are
narrower in the classical limit. Another clear feature of this
nonlinear system revealed by $S(q,\omega)$ is that the breather bands
are irregularly spaced.

To summarize, we have demonstrated that quantum breathers exist as
eigenstates of the simple nonlinear lattice system studied with a
renormalized mass corresponding to a narrow but finite bandwidth.  The
quantum breathers have been identified as robust nonlinear solutions
with their own characteristic spatial localization and multi-frequency
dynamics.  Finally, we predict that the dynamic structure factor is
strongly enhanced at momentum and energy transfers corresponding to
breather excitations and correspondingly reduced elsewhere.  This novel
result should apply in more general situations and provide exciting new
possibilities for the identification of breathers and other nonlinear
excitations by neutron scattering.  The numerical approach used here is
readily applied to other nonlinear lattices and to electron-phonon
coupled models, for which results will be presented elsewhere.

We thank D. Cai, S. R. White, and J. Zang for useful discussions,
and E. Y. Loh, Jr. of Thinking Machines Corp. for helpful discussion.
This work was performed at the Advanced Computing Laboratory of Los Alamos
National Laboratory under the auspices of the U.S. Department of
Energy. M.I.S.\ was also supported in part by  Natural Sciences and Engineering
Research Council of Canada, and the Ontario Center for Materials Research.


\begin{references}
\bibitem{einstein} The Hamiltonian (\ref{eq1}) can be diagonalized
using different bases. We chose an Einstein basis since it is simple
computationally.
\bibitem{h-footnote}
Rather than using the Davidson algorithm to calculate
the first $\alpha$$\sim$200 wavefunctions, we target
several energies $\lambda_k$,  $\epsilon_0$$<$$\lambda_k$$<$$\epsilon_\alpha$.
For each $\lambda_k$ we calculate the lowest $\sim$20 eigenstates of
$(H-\lambda_k)^2$, using enough $\lambda_k$ to fill in the spectrum.
This is computationally more efficient,
yields the excited eigenstates within
different energy windows, and prevents loss of states in the windows.
We developed parallel algorithms on a CM-5 machine with a 2,048
processor partition, and were able to find the first few hundred
eigenstates of a multi-million matrix in a few hours.
\bibitem{freq-note}
Most of the classical or semi-classical
breather solutions which have been studied
have a single frequency or commensurate relationship between internal
frequencies in the low amplitude weakly nonlinear regime.
However, incommensurate multi-frequency classical breathers
are also known in discrete models (Ref.~\cite{cai,takeno}).
\end{references}
\end{document}